\documentclass[twocolumn]{aastex631}

\usepackage{amsmath}
\usepackage{xcolor}
\usepackage{mathtools}
\usepackage{soul}
\usepackage{txfonts}

\begin{document}

\title{First observation of double-peaked O\,{\sc i} emission in the near-infrared spectrum of an active galaxy}

\author[0000-0003-4153-4829]{Denimara Dias dos Santos}
\affiliation{Instituto Nacional de Pesquisas Espaciais (INPE), Av. dos Astronautas, 1.758, Jardim da Granja, 12227-010, São José dos Campos, São Paulo, Brazil}
\affiliation{Istituto Nazionale di Astrofisica(INAF), Osservatorio Astronomico di Padova, Vicolo dell’ Osservatorio 5, IT35122 Padova, Italia}

\author[0000-0002-7608-6109]{Alberto Rodríguez-Ardila}
\affiliation{Laboratório Nacional de Astrofísica (LNA), MCTI, R. dos Estados Unidos, 154, Nações, 37504-364, Itajubá, Minas Gerais, Brazil}
\affiliation{Instituto Nacional de Pesquisas Espaciais (INPE), Av. dos Astronautas, 1.758, Jardim da Granja, 12227-010, São José dos Campos, São Paulo, Brazil}

\author[0000-0002-5854-7426]{Swayamtrupta Panda}
\affiliation{Laboratório Nacional de Astrofísica (LNA), MCTI, R. dos Estados Unidos, 154, Nações, 37504-364, Itajubá, Minas Gerais, Brazil}

\author[0000-0001-9719-4523]{Murilo Marinello}
\affiliation{Laboratório Nacional de Astrofísica (LNA), MCTI, R. dos Estados Unidos, 154, Nações, 37504-364, Itajubá, Minas Gerais, Brazil}







\begin{abstract}
Double-peaked profiles associated with the broad-line region (BLR) of active galactic nuclei (AGNs) are regarded as the clearest evidence of the presence of an accretion disk. They are most commonly detected by means of optical spectroscopy in the Balmer lines and in the Mg\,{\sc ii}\,$\lambda$2798 ultraviolet line. Here, we report the first unambiguous detection of a double-peak broad emission line associated with the O\,{\sc i}~$\lambda$11297 emission line in the near-infrared (NIR) in the local Seyfert~1 galaxy III~Zw~002. 
Additionally, we detect simultaneously in the spectrum the double-peak emission in the Pa$\alpha$ line and very likely in the He\,{\sc i}~$\lambda$10830. This is the first time that several broad double-peaked NIR emission lines have been detected simultaneously.
The double-peaked profiles are fit using a disk-based model, with an additional Gaussian component attributed to nondisk clouds, which represents the classical BLR. Our results obtained from the fits reveal important parameters, such as disk inclination and geometry. From the double-peaked profile fits, we suggest that the BLR in III\,Zw\,002 has a disk-like geometry, as it extends up to the outer edge of the BLR.

\end{abstract}



\section{Introduction} \label{sec:intro}

Active galactic nuclei (AGNs) are primarily powered by a supermassive black hole (SMBH) lying at the center of a host galaxy that is accreting matter from its vicinity. The radiation from the central source ionizes the inner dense gas, the broad-line region (BLR). The BLR likely extends up to the outer region of the accretion disk, which is delimited by the dust sublimation radius or the onset of the dusty molecular torus \citep{Nenkova2008, Kishimoto2011TheInterferometer, Koshida2014,Panda2020OpticalModelling}. Studies using the reverberation mapping (RM) technique for over 100 AGNs have estimated the BLR line-emitting radius (especially for the H$\beta$ line) around light days to a few light weeks, confirming the compact nature of this region \citep{Blandford1982REVERBERATIONQUASARS, Kaspi2000ReverberationNuclei, Peterson2004ReverberationNuclei, Bentz2009THEAGNs, Du2015SUPERMASSIVEACCRETION,Du2018Mapping, Grier2013TheMaps, Grier2017TheCampaign}.

AGN spectra typically exhibit symmetric or asymmetrical BLR emission line profiles, with the latter displaying blue or red asymmetries, most commonly attributed to outflows or inflows, respectively \citep{Du2018Mapping, Bao_etal_2022}. However, a fraction of AGNs shows complex features, departing from the signatures expected when gas in the BLR is either in outflow or inflow. Among these interesting sources are the ones demonstrating the double-peaked broad emission lines. They are associated with a flattened disk-like BLR structure, first proposed by \cite{Chen1989Kinematic102B} and \cite{Chen1989STRUCTURE102B} in the optical region after observing the H$\alpha$ line in the Type-I AGN Arp~102B. Later, several works in the literature suggested that the double-peaked emission originates from the outer parts of a relativistic Keplerian disk of gas of about $\sim$\,1000 gravitational radii\footnote{$\xi$\,=\,R$_{g}$\,=\,G\,M$_{\rm BH}$/c$^{2}$, where M$_{\rm BH}$ is the black hole mass.} surrounding the SMBH \citep{Eracleous1995EllipticalNuclei,Eracleous2003CompletionNuclei, strateva, Schimoia2017Evolution7213, Storchi-Bergmann2017Double-PeakedNuclei}.
Other works have suggested alternative explanations for the double-peaked emission, such as hybrid disk models, a binary black hole, or a BLR dominated by a radiatively accelerated wind (see \cite{Eracleous2009Double-peakedNuclei} for a complete review), but the disk-like geometry for BLR still has the best-supporting evidence from theoretical and observational grounds \citep{Eracleous2003CompletionNuclei, Schimoia2017Evolution7213, Storchi-Bergmann2017Double-PeakedNuclei}.

Double-peaked emission lines are most frequently observed in optical spectra, particularly in the  Balmer lines. In the UV region, evidence of the detection of double-peak emission has been reported in the Mg\,{\sc ii} line \citep{strateva}. 
However, in the near-infrared (NIR), reports of the observation of double-peak line profiles are very scarce. To the best of our knowledge, only one object, SDSSJ153636.22+044127.0 \citep{zhang2019}, presents evidence of double-peak BLR lines. Pa$\beta$ and He\,{\sc i}\,$\lambda$10830 were fitted using the disk model commonly employed in the optical. However, in general, other well-known double-peaked sources in the optical, such as NCG1097 \citep{2002reunanen, Storchi-Bergmann2017Double-PeakedNuclei},  do not show double-peaked profiles in the NIR.

III~Zw\,002, also named Mrk~1501 and PG~0007+106, is a Seyfert~1 AGN classified as a radio-intermediate source at z\,$=$\,0.089. 
Studies over the years have reported strong variability in III~Zw\,002 from high energies ($\gamma$-ray and X-ray), to radio \citep{Brunthaler2000IIIGalaxy, Brunthaler2005TheIIIZw2:,Liao2016DiscoveryVariability,Gonzalez2017The2}. These works have provided information regarding the central engine and jet activities on the subparsec scale for this source.

In the optical the first evidence of variability as well as the presence of an asymmetric profile in the H$\beta$ line in III\,Zw\,002 was noticed by \citet{Popovic2003TheVariability}. They constructed a mean and an RMS spectrum using 41 spectra gathered at different epochs spanning between 1972 and 1998 in the wavelength region from 4740 to 5900~\AA. They saw a persistent feature/bump in the blue and red shoulders of the full profile of H$\beta$, fitted using a disk model plus a Gaussian component.

\citet{Grier2012ReverberationGalaxies} showed, from a dense reverberation mapping campaign with about 80 spectra taken between 2010 January and 2011 July, a similar result as that of \citet{Popovic2003TheVariability}. They found a persistent feature $\sim$30-40\AA\ redward of the line center of H$\beta$, in both their mean and rms spectrum. However, the blueward feature, $\sim$35\AA\ of the line center, showed up only in their mean spectrum, unlike the result from \citet{Popovic2003TheVariability}. \citet{Grier2012ReverberationGalaxies} also showed the velocity-resolved maps around the H$\beta$ region that demonstrate an extended shoulder redward of the central wavelength that corresponds to a time lag of $\sim$2-10 days, whereas the line center corresponds to a time lag of 15.5$^{+2.2}_{-1.9}$ days, resulting in a virial black hole mass of $M_{\rm BH}$ = 3.34$\pm$0.49$\times 10^7 M_{\odot}$. The black hole mass estimated for this source using the $M_{\rm BH}$ - $\sigma_{\star}$ relation is about 5.5 times more massive, i.e., (1.84$\pm$0.27)$\times 10^8 M_{\odot}$ reported in the same work \citep{Grier2012ReverberationGalaxies}.

In this work, we carry out a spectroscopic analysis in the NIR of the Seyfert~1 galaxy III\,Zw\,002 aimed at confirming the double-peaked nature of this source. The large wavelength coverage of the spectrum (0.9 $-$ 2.4~$\mu$m) will allow us to study simultaneously several broad-line features emitted by the BLR  (O\,{\sc i}, H\,{\sc i}, and He\,{\sc i}). We highlight that this is the first time that such an analysis is made on this source covering the $JHK$ bands, allowing us to explore a new wavelength interval in a yet very interesting source. 
This Letter is organized as follows: In Section \ref{sec:data} we present the data and observations of the NIR spectrum of III~Zw~002. The analysis of the NIR spectrum, including the detection of double-peaked broad permitted lines, is in Section \ref{sec:model}.  Results from the line profile fitting to disk models are presented in Section \ref{sec:results}. A discussion of the results is in Section \ref{sec:discussion}. The main conclusions are in Section \ref{sec:conclusions}. Throughout the paper, we use a flat $\Lambda$CDM cosmology with $H_{\rm 0}$ = 70~km~s$^{-1}$~Mpc$^{-1}$, $\Omega_{\rm m}$ = 0.30, and $\Omega_{\rm vac}$ = 0.70.

\section{Observations and data reduction} \label{sec:data}

\begin{figure*}[!htb]
\centering	
\includegraphics[width=\textwidth]{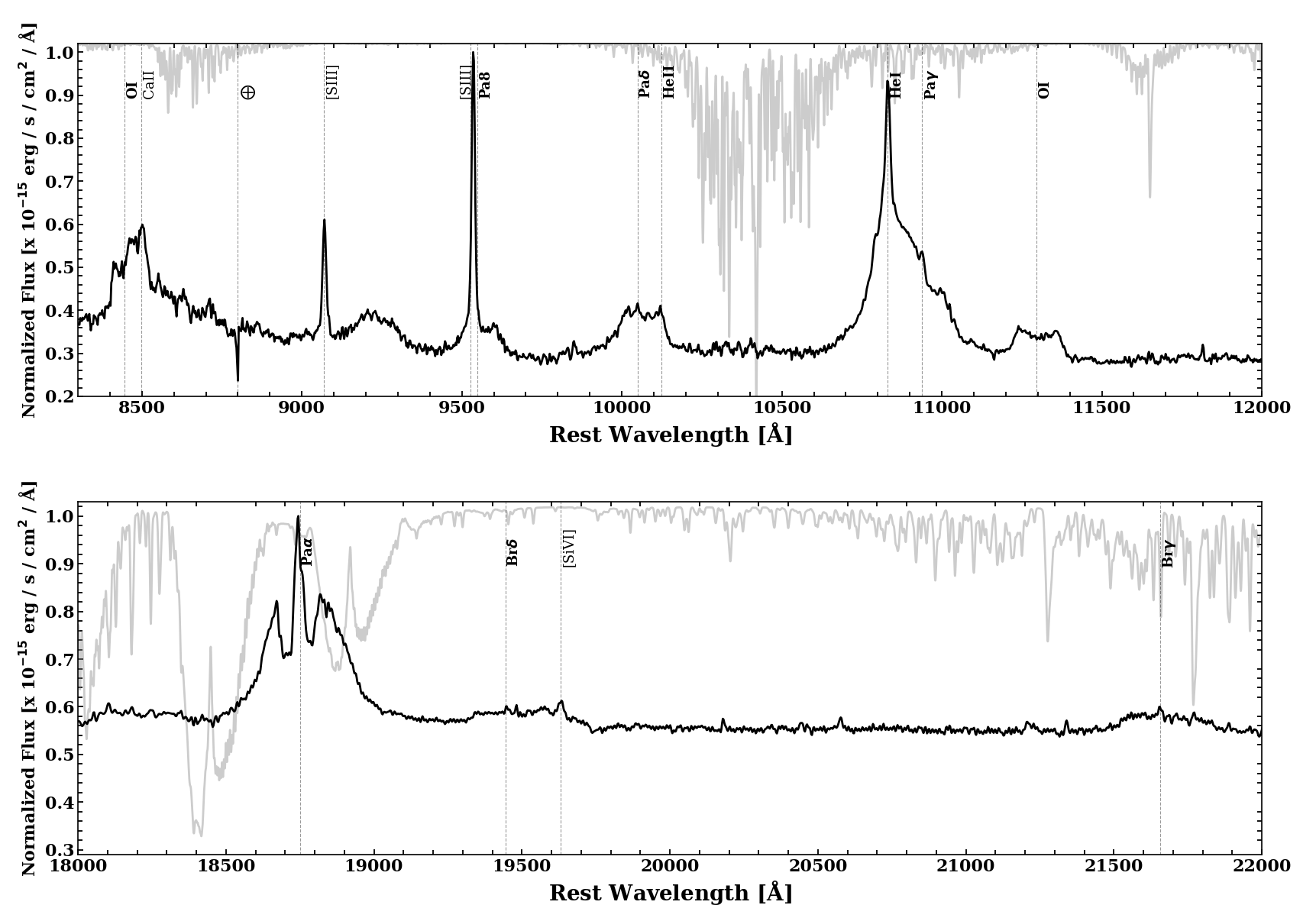}
\caption{NIR spectra of the III~Zw~002 in rest wavelength, observed with GEMINI-N/GNIRS on 2021-07-28. Here we show the spectral region of J and K bands, respectively covering 0.7$\rm\mu$m$-$1.3$\rm\mu$m (upper panel), and 1.7$\rm\mu$m$-$2.3$\rm\mu$m (down panel). We also mark the most notable emission lines, and in bold-face the emission lines exhibiting double-peaked feature. The background atmospheric transmission filter is represented in gray. 
}
 \label{figure1}
\end{figure*}

NIR spectra of III\,Zw\,002 were collected on the night of 2021 July 28 at the Gemini North 8.1\,m telescope using the GNIRS spectrograph in the cross-dispersed mode. This configuration covers simultaneously the 0.8\,$-$\,2.5 $\mu$m spectral range. A slit width of 0.675'' $\times$ 7'' and a 32~l/mm grating were employed, providing a spectral resolution of 1700. Data were acquired by nodding along the slit using the ABBA pattern. Eight individual exposures of 150~s each were collected. The telluric star A0V HIP186 (V=6.3 mag) was observed right after the target at an airmass of 1.052.  

The observations were reduced using the nonofficial pipeline {\sc xdgnirs} (v\,2.0), with the basic scripting set up by \citet{Gonzalez-Martin2009DustPipeline} and fully tested and documented by \citet{Mason2015THEGALAXIES}. 
{\sc xdgnirs} delivers a fully reduced, wavelength- and flux-calibrated 1D spectrum with all orders combined  \citep{Mason2015THEGALAXIES}. Briefly, it cleans the 2D images from radiative events and prepares a master flat to remove pixel-to-pixel variations. Thereafter, the s-distortion solution is obtained from daytime pinhole flats and applied to the science and telluric images to rectify them. Arc frames were used to find the wavelength dispersion solution, followed by the extraction of 1D spectra from the combined individual exposures. The telluric features from the science spectrum were removed using the spectrum of the telluric standard. Thereafter, the flux calibration is achieved assuming a blackbody shape for the standard star \citep{pecaut2013intrinsic} scaled to its $K$-band magnitude \citep{skrutskie2006two}. Finally, the different orders were combined in a single 1D spectrum. The final spectrum in rest-frame wavelength is shown in Figure~\ref{figure1} except the part corresponding to the $H$ band because it has no spectral features of interest to this work.

\begin{figure}[!htb]
\centering	
\includegraphics[width=\columnwidth]{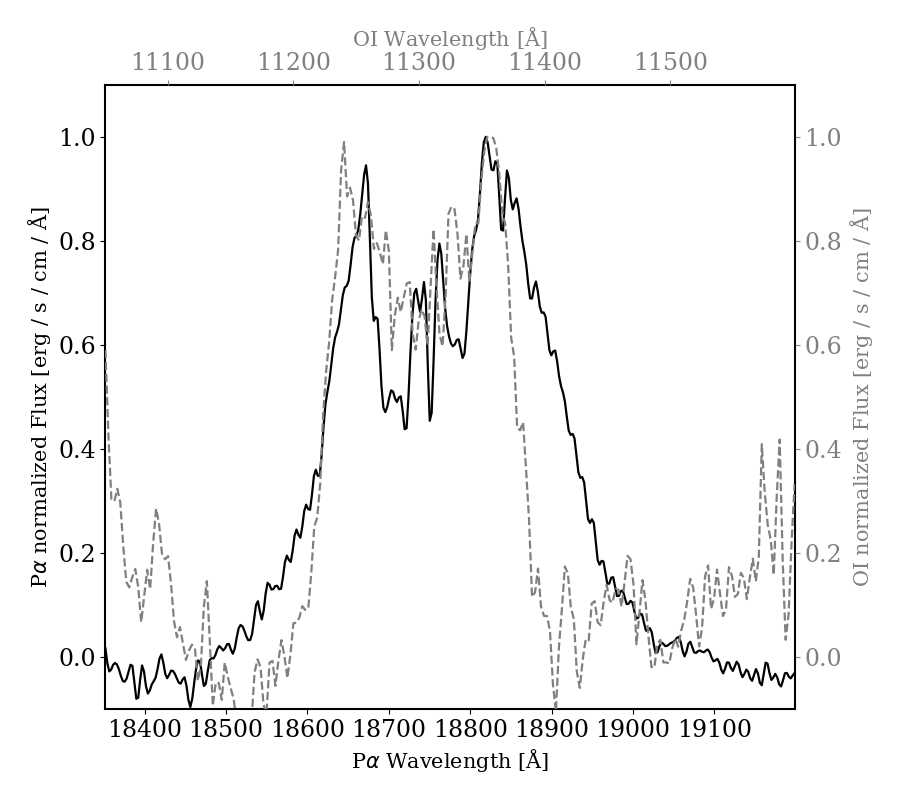}
\caption{Overlapping double-peak profile for Paschen $\alpha$, in the solid black line, and O\,{\sc i} in the dotted gray line. We subtract the narrow component of Paschen $\alpha$ to compare the profiles.}
\label{figure2}
\end{figure}

The NIR spectrum of III~Zw~002 displays the typical features of a Type-I AGN, including broad H\,{\sc i} and He\,{\sc i} lines \citep{Riffel2006}. Pa$\beta$ is not detected, as it falls in the gap between the $J-$ and $H-$ bands. An inspection of Fig.~\ref{figure1} reveals unambiguously double-peaked profiles in the lines of Pa$\alpha$ and O\,{\sc i}~$\lambda$11287 (hereafter O\,{\sc i}). Figure~\ref{figure2} compares the profiles of the latter two lines, normalized to unity. Although both profiles display conspicuous differences mostly in the degree of asymmetry and width, their double-peak structure is remarkable.
To the best of our knowledge, this is the first time in the literature that such a profile is reported in both lines. Moreover, in Pa$\alpha$, the base of the line exhibits extended wings, not seen in O\,{\sc i}. 
Unlike our observations in the NIR, evidence of a double-peaked profile in the optical region, in particular in the Balmer lines, is rather subtle as the profiles do not exhibit the characteristic ``horn'' feature.  

It can be argued that the detection of the double-peak profiles in the NIR is due to residuals left after division by the telluric star. This is because the spectral region where both lines are located can be strongly affected by telluric features depending on the redshift of the target. However, this is not the case in III\,Zw\,002, as both lines fall in regions less prone to be telluric affected. The atmospheric transmission spectrum in the GNIRS data is plotted in light gray in Fig.~\ref{figure1} along with the observed spectrum. It can be seen that O\,{\sc i} is located in a region free of telluric features. Thus, from the O\,{\sc i}  point of view, we can unambiguously state that the double-peak detection is real.
In Pa$\alpha$, the red wing of that line falls very close to a strong H$_2$O absorption. However, the blue peak as well as the start of the red peak of the line are safe from these effects. 

It is also important to notice that the air mass during the observation of III\,Zw\,002 varied very little (1.011$-$1.143). In addition, the difference in airmass between the telluric standard and the galaxy was merely 0.009. The goodness of the telluric correction can be clearly appreciated in Fig.~\ref{figure1}, as the residuals at the regions of bad atmospheric transmission are very close to the RMS of the continuum at the locations free of telluric features. Moreover, the reduction process was done by three different collaborators, in an independent way, to cross-check the final spectrum. In all cases, the double-peaked profiles were very clear, with an excellent match in shape and flux, showing differences of less than 5\%. 

Finally, after a visual inspection and analysis of the NIR spectrum, in addition to O\,{\sc i} and Pa$\alpha$, we found compelling evidence of double-peaked profiles in the He\,{\sc i}~$\lambda$10831, Pa$\delta$, and Br\,$\delta$ lines. 
However, these lines are highly blended or have a smaller signal-to-noise ratio (S/N) than O\,{\sc i} and Pa$\alpha$ making their modeling complex. For that reason, we will conduct a comprehensive analysis of them in a forthcoming extended paper. In this Letter, we focus on the well-isolated lines with a great S/N to apply the disk model.

\section{Double-peaked emission spectral fitting} \label{sec:model}

In order to explain the nature of the double-peaked profiles in III\,Zw\,002, we assume that at least part of the emission originates from different portions of a circular Keplerian disk.

In order to confirm this hypothesis, the disk model of \citet{Chen1989Kinematic102B} was employed. It has already been used to explain the observed double-peaked profiles in NGC\,1097 and 3C\,390.3. 
The disk model consists  of eleven free parameters: the inner ($\xi_{1}$) and external ($\xi_{2}$) radii, the disk inclination $i$ relative to the sky plane, and parameters related to disk spiral arms. 
The total line flux of the emission line is given by the integration of the specific intensity from each location in the disk, i.e., in the frame of an emitting particle, in azimuthal $\phi-\Psi_{0}$ and radius distance. The emissivity is modeled by the law described by the q$_{1}$ index. When it hits the saturation radius – or maximum emissivity ($\xi_{q}$)$-$the index of the emissivity law changes to q$_{2}$. 
Moreover, the disk model also includes spiral arms, which may exhibit brightness contrast ($A$) between them. In addition, the arms change the emissivity according to the azimuthal distance $\phi-\Psi_{0}$ from the ridge line to both sides of the arm. It is assumed that the arm's emissivity behaves as a Gaussian function with an azimuthal width parameter, $\delta$.
The set of equations that govern the disk emission and applied to the case of III\,Zw\,002 was obtained from  \citet{Chen1989STRUCTURE102B, Chen1989Kinematic102B, Eracleous1995EllipticalNuclei, Eracleous2003CompletionNuclei, Schimoia2012SHORT1097, Schimoia2017Evolution7213} and \citet{Storchi-Bergmann2017Double-PeakedNuclei}.

In the fitting process, we also included a Gaussian component associated with clouds outside the disk structure, likely located outside the disk plane. The Gaussian component contributes to the central hump in the line profile, with a typical FWHM larger than 2000 km\, s$^{-1}$. 
This component was shown to be critical to explain the ``boxy'' profile observed in O\,{\sc i} and the extended wings in Pa$\alpha$. It is very similar to the one recently reported by \cite{HungDouble-peakedEvent} in the H$\alpha$ line. 

Once we obtained the best parameters of the disk model, we added the Gaussian component. Thereafter, we applied a scale factor to  find the best contribution of both components to the total profile. To this purpose, we use {\sc lmfit} \citep{Newville2014LMFIT:Python}, a {\sc python} library that uses the Levenberg$-$Marquardt (\textit{leastsq}) method. 

\begin{table}[]\scriptsize
\begin{center}
\begin{tabular}{ccc}
\hline
Parameter                            & O\,{\sc i}                 & Pa$\alpha$         \\ \hline
q$_{1}$                              & -1                 & -1                 \\
$\phi_{0}$ (º)                          & 151                & 165                \\
i (º)                                   & 18                 & 18                 \\
$\delta$ (º)                            & 50                 & 18                 \\
p (º)                                   & 176                & 176                \\
A                                    & 0.5                & 3.5                \\
$q_{2}$                              & 2.5                & 2.5                \\
$\xi_{1}$ (R$_{\rm g}$)                            & 800 (8.38 lt-day)               & 800 (8.38 lt-day)               \\
$\xi_{q}$ (R$_{\rm g}$)                             & 1800 (18.86 lt-day)              & 1600 (16.77 lt-day)               \\
$\xi_{2}$ (R$_{\rm g}$)                             & 5000 (52.43 lt-day)              & 5000 (52.43 lt-day)              \\
$\sigma$ (km s$^{-1}$)                        & 1175               & 1175               \\
Factor                                & 0.72\,$\pm$\,0.03      & 0.37 \,$\pm$\, 0.01      \\ \hline
\multicolumn{3}{c}{Gaussian Component}                                         \\ \hline
Peak $^{*}$ (BC)                            & 0.17\,$\pm$\,0.01      & 0.18 \,$\pm$\, 0.011      \\
Center (BC, in $\AA$)                          & 11303.92\,$\pm$\,2.28  & 18787.53 \,$\pm$\, 1.76 \\
FWHM (BC, in km s$^{-1}$)                            & 2474.56\,$\pm$\,387.25 & 5460.85 \,$\pm$\, 150.49 \\
Peak $^{*}$ (NC)                            &        ...           & 0.53 \,$\pm$\, 0.02      \\
Center (NC, in $\AA$)                          &      ...              & 18745.17 \,$\pm$\, 0.25  \\
FWHM (NC, in km s$^{-1}$)                            &           ...         & 410.91 \,$\pm$\, 10.76   \\
Normalized flux$^{*}$   & $\times$6.92$\times10^{-17}$   & $\times$2.03$\times10^{-16}$   \\ \hline
\end{tabular}
    \scriptsize\item Rg\,=\,0.010~lt-day, assuming the black hole mass of \,1.84$\times$10$^{8}\,$M$\odot$ \citep{Grier2012ReverberationGalaxies}.
    \scriptsize\item $^{*}$ [erg s$^{-1}$ cm$^{-2}$ \AA$^{-1}$]
\end{center}\caption{Fitting parameters for the Disk+BLR model for the OI and Pa$\alpha$ Lines.}
\end{table}\label{tabela1}

\section{Analysis and results}\label{sec:results}
\subsection{Results from the spectral fitting}

\begin{figure*}[!htb]
\centering
\includegraphics[width=\textwidth]{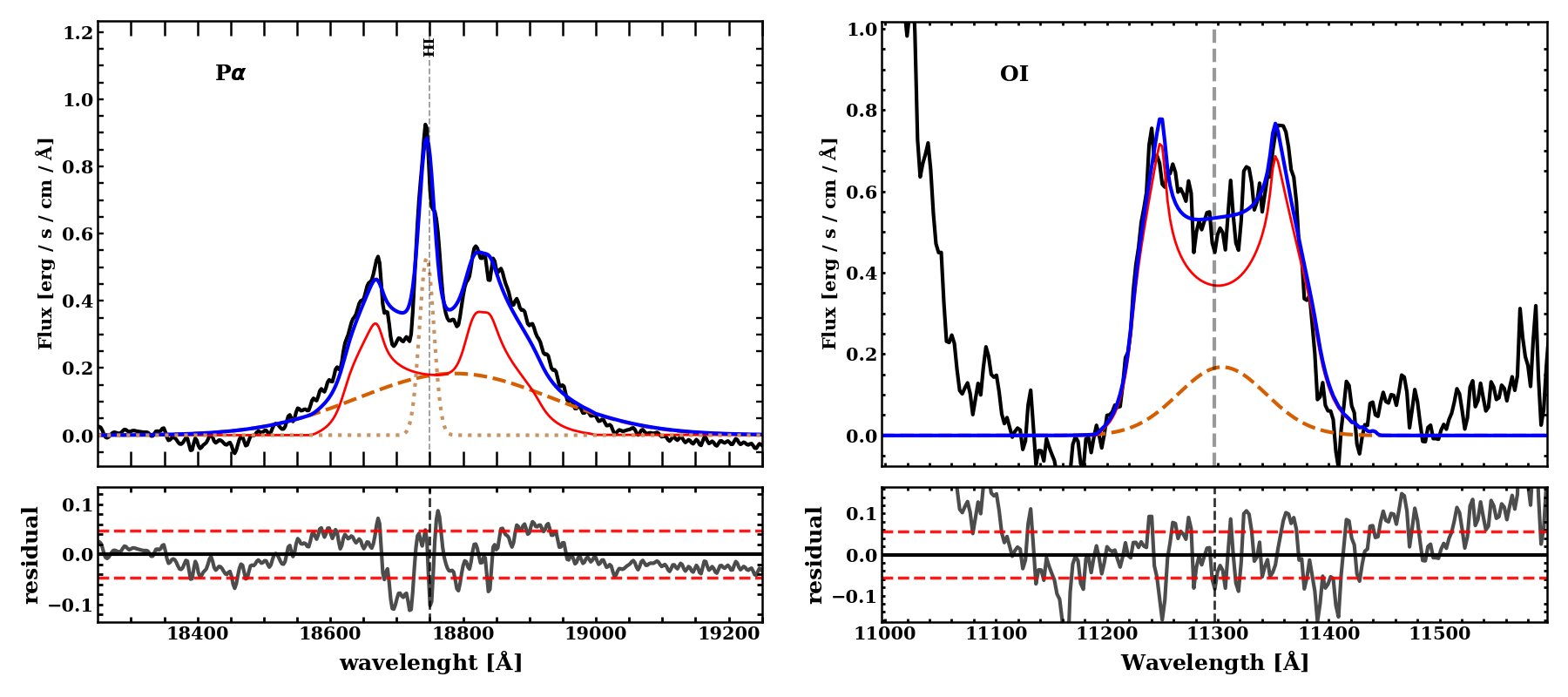}
\caption{Pa$\alpha$ and O\,{\sc i} emission line fits. Left: observed profile for Pa$\alpha$ (black solid line), fitted double-peaked component (red solid line), and residuals in gray line. The dotted and dashed red lines are narrow and broad Gaussian components, respectively. Right: fit of the O\,{\sc i}, where the colors have the same meaning for the disk and broad Gaussian component.}
\label{figure3}
\end{figure*}

Table~\ref{tabela1} shows the parameters obtained for the disk model and Gaussian components. The Pa$\alpha$ line presents well-separated peaks (see figure~\ref{figure3}). The $A$ parameter, the contrast in the spiral arms, is responsible for reproducing the peak intensities. Here, we found a value of $A=$\,3.5. The saturation radius ($\xi_{q}$) found from the fit for Pa$\alpha$ is $\sim$1600 gravitational radii. This parameter is responsible for fitting the prominence of the profile wings and is an essential parameter related to the maximum emissivity of the disk spiral arms. The parameter ``factor'' in Table~\ref{tabela1} is a scale factor applied to the disk model regarding the total line profile. From the fit we obtained the values of 0.72\,$\pm$\,0.07 and 0.37\,$\pm$\,0.01 for O\,{\sc i} and Pa$\alpha$, respectively.

We note that O\,{\sc i} exhibits peaks with similar intensities, resembling a “boxy'' shape, unlike the Pa$\alpha$ line profile, as we may see in Figure~\ref{figure3}. The parameter $A$, obtained from the fit result for O\,{\sc i}, shows the value 0.5, which is responsible for reproducing the slight difference between the intensities of blue and red peaks. The parameters associated with the disk, such as internal and external radius, pitch angle, disk inclination, and emissivity power-law index are equal for both lines, confirming the reliability of our model. In contrast, we obtained different values in both lines for the parameters related to the spiral arms, such as the saturated radius, azimuthal position, and spiral arms distance. We found, for instance, a value for $\xi_{q}$ of about 1800 gravitational radii, i.e., larger than that of Pa$\alpha$. This is in agreement with the literature, which suggests that the O\,{\sc i} line is likely produced in the outer BLR \citep{LoliMartinez-Aldama2015, Marinello2016, Panda2020OpticalModelling, 2021POBeo.100..333P}.

Regarding the Gaussian components, to fit the Pa$\alpha$ profile, we added two components: a broad one (BC) associated with the nondisk BLR region contributing to fit profile wings and a narrow component (NC) associated with the emission produced in the narrow line region (NLR). O\,{\sc i} does not display an NC, as the observed emission is produced in its entirety in the BLR \citep{RodriguezArdila2000VisibleGalaxies, Matsuoka2008LowIonizationLines}. 

\section{Discussion}
\label{sec:discussion}

We have successfully modeled in III\,Zw\,002 the double-peaked profile detected in O\,{\sc i} and Pa$\alpha$ assuming a Keplerian disk model. The physical interpretation of this result points out that both emission lines originate in a BLR distribution of clouds dominated by a planar geometry, with orbits well described by those of a Keplerian disk. Interestingly, from the modeling, the disk is geometrically more extended in O\,{\sc i} than in Pa$\alpha$. This result is consistent with several works that have  demonstrated through kinematic and photoionization models that the O\,{\sc i} line originates in the outer portion of the BLR  \citep{LoliMartinez-Aldama2015, Marinello2016, Marinello2020Panchromatic1092, Panda2020OpticalModelling, 2021POBeo.100..333P}. 

Our results present solid evidence that the bulk of the BLR follows a planar geometry, similar to several other papers suggesting that other AGNs also have a disk-like BLR \citep{pancoast14, Storchi-Bergmann2017Double-PeakedNuclei, sergeev20}. The consistency between the model fits made to Pa$\alpha$ and O\,{\sc i} points out that both lines are emitted by the same disk, providing further support to our hypothesis.

In general, double-peaked emission lines have been observed in a single line in AGN spectra \citep{Eracleous2003CompletionNuclei, strateva, Schimoia2017Evolution7213, Storchi-Bergmann2017Double-PeakedNuclei}. Here, we simultaneously found double-peaked emission in two different BLR lines. Furthermore, from the disk parameters obtained in the fit, Pa$\alpha$ and O\,{\sc i} are emitted in different locations in the disk, i.e., different radii and azimuthal distances. That would explain the different shapes observed in both profiles, even though the same disk emits them. The Pa$\alpha$ originates in a complex internal region, nearer to the AGN than O\,{\sc i}. That makes the former line suitable to map the variability of the AGN continuum emission. In contrast, otherwise, O\,{\sc i} likely would not change considerably with time because of its larger distance from the central ionizing source.

From the reverberation campaigns carried out on III\,Zw\,002, \citet{Grier2012ReverberationGalaxies, Grier2013TheMaps, Grier2017TheCampaign} constructed velocity-delay maps by means of the measured time delays in different velocity bins of the H$\beta$ emission line. The authors gathered evidence of (i) an ``extended BLR” (see Figure~4 \cite{Grier2013TheMaps}); (ii) radial stratification of the gas ionization; and (iii) the geometry of BLR consistent with a thick inclined disk.   
Due to the employment of different spectral lines, i.e., H$\beta$ by \citet{Grier2017TheCampaign}, and Pa$\alpha$ by our study, we expected a minor variation in the physical scale. Specifically, \citet{Grier2017TheCampaign} reported a maximum emissivity radius of around 15.5~ld-day for H$\beta$, while our findings for the Pa$\alpha$ line indicate 16.77~ld-day. The findings of \citet{Grier2017TheCampaign} align with our conclusions regarding the BLR disk-like geometry of III\,Zw\,002, providing further validation and support for the results presented in this Letter. 
To the best of our knowledge, this is the first time that several broad double-peaked NIR emission lines have been detected simultaneously in an AGN, opening up the possibility of studying the structure of the BLR using multiwavelength observations, including several broad double-peaked emission lines.

Here, it is important to remember that the disk model considers spiral arms, which may exhibit brightness contrast between the arms and change in emissivity according to the distance from the AGN \citep{Chen1989STRUCTURE102B}. These BLR spiral arms are composed of matter that acts in the dragging process under the SMBH gravitational potential \citep{Shapovalova2008Long-termCorrelations, Shapovalova2010Long-termProfiles, Grier2012ReverberationGalaxies, Grier2013TheMaps, Grier2017TheCampaign, Shapovalova2012}. That process may be responsible for the asymmetries in the line profile observed in O\,{\sc i} and Pa$\alpha$.

The BH mass of III\,Zw\,002 can be estimated assuming that the gas motion within the BLR is dominated by the gravitational potential of the BH and using the virial equation $M_{\rm BH}=f\times R_{BLR}\times\Delta V^2/G$, where $G$ is the gravitational constant, $\Delta V$ is the velocity dispersion of the clouds (given by the FWHM of H$\beta$), $R_{BLR}$ is the BLR radius, and $f$ is the virial factor \citep{Peterson2004BlackMeasurements}.
Despite the virial factor being a key element in estimating the black hole masses, it directly depends on the disk inclination $i$, which cannot be measured directly from observations \citep{Peterson2004BlackMeasurements, Bentz2009THEAGNs, Du2015SUPERMASSIVEACCRETION,  Grier2012ReverberationGalaxies, Grier2013TheMaps, Grier2017TheCampaign, Du2019TheNuclei, Pandaetal2019, Dallabont2022A}.

Thus, the virial factor carries an intrinsic bias into the estimative of black hole masses. By means of the disk fitting, the disk inclination is derived among the fitted parameters and the virial factor can be estimated by $f=1/(\kappa^2 + sin^2 i)$, where $\kappa$ carries information on the disk geometry. Using the inclination we estimated ($i$=18$^{\rm o}$) and four different values for $\kappa$: 0.3, 0.2, 0.1, and 0.0 (where $\kappa$=0 refers to a flat disk geometry) \citep{ Collin2006SystematicInclination, Grier2013TheMaps, Storchi-Bergmann2017Double-PeakedNuclei, Pandaetal2019}, we obtained virial factors 5.39, 7.38, 9.47, and 10.47, respectively. These results are consistent with the inclinations measured for NGC\,5273 ($i$=16$^{\rm o}$), NGC\,3227 ($i$=17$^{\rm o}$), and NGC\,5548 ($i$=19$^{\rm o}$), and with estimated virial factors of 12, 11.7, and 9.4, respectively \citep{Storchi-Bergmann2017Double-PeakedNuclei}.

Moreover, the Pa$\alpha$ FWHM value (5460.85$\pm$150.49 km s$^{-1}$) we measure in this letter is close to the FWHM value obtained from the rms spectrum for H$\beta$ (5054$\pm$145 km s$^{-1}$) from \citet{Grier2012ReverberationGalaxies}. Thus, we employed the FHWM of Pa$\alpha$ instead of that of H$\beta$ to estimate the BH mass, since the measured FWHM of both lines are consistent.  
Furthermore, we used the same BLR radius obtained by the later authors.
Using these parameters, we estimate BH masses of $4.86\times10^8$$M_{\odot}$,  $6.65\times10^8$$M_{\odot}$, $8.55\times10^8$$M_{\odot}$, and  $9.44\times10^8$$M_{\odot}$, which are in good agreement with BH mass estimated by RM, $1.84\times10^8$$M_{\odot}$, using $f$=5.5 \citep{Grier2012ReverberationGalaxies}.

The results above fall within the relation presented by \citet{Storchi-Bergmann2017Double-PeakedNuclei} (see Figure 8 in their paper), which shows that low inclinations (implying larger virial factors) are associated with larger FWHM of broad lines. Our results allowed us to set an important constraint on the $f$ factor and a more robust estimation of the BH mass. 
In this context, the disk modeling approach can be a powerful alternative to minimize the uncertainties related to the virial factor, which is an essential parameter to probe the BLR geometry and cloud distribution \citep{Collin2006SystematicInclination, Storchi-Bergmann2017Double-PeakedNuclei, Pandaetal2019}.

\section{Conclusions}\label{sec:conclusions}

We report, for the first time, the simultaneous detection of a double-peaked profile in the O\,{\sc i}~$\lambda$11297 and Pa$\alpha$ emission lines in the local Seyfert~1 galaxy III\,Zw\,002. The analysis presented in this Letter employs the NIR spectrum for this source covering the $JHK$ bands for the first time, allowing us to explore double-peaked profiles in a new wavelength interval.

The O\,{\sc i}~$\lambda$11297 emission line is located in a region free of atmospheric absorptions,  making our detection irrefutable and confirming the double-peaked nature of that source.  As a result of the successfully applied disk model to the O\,{\sc i} emission profile, we can infer the geometry of the outer region of the BLR. From the modeled evidence, we suggest for III~Zw~002 that the O\,{\sc i} emission comes from a flattened, low-ionization line-emitting region.

From the model fitting parameters, we also find a low inclination angle for the disk, $i=$18$^{\rm o}$. In this context, the disk modeling approach can be a powerful alternative for minimizing uncertainties related to the virial factor, which is an essential parameter to probe the BLR geometry and cloud distribution, as well as black hole mass estimation in the AGNs.

\begin{acknowledgments}
The authors gratefully receive financial support through the Brazilian Agencies: Agency of Coordenação de Aperfeiçoamento de Pessoal de Nível Superior (CAPES), and Conselho Nacional de Desenvolvimento Cient\'{\i}fico e Tecnol\'ogico (CNPq). The first author was financed in part by the CAPES – Finance Code 001.
\end{acknowledgments}

%

\facilities{Gemini North (GNIRS).}

\bibliography{references}{}

\begin{thebibliography}{}
\expandafter\ifx\csname natexlab\endcsname\relax\def\natexlab#1{#1}\fi
\providecommand{\url}[1]{\href{#1}{#1}}
\providecommand{\dodoi}[1]{doi:~\href{http://doi.org/#1}{\nolinkurl{#1}}}
\providecommand{\doeprint}[1]{\href{http://ascl.net/#1}{\nolinkurl{http://ascl.net/#1}}}
\providecommand{\doarXiv}[1]{\href{https://arxiv.org/abs/#1}{\nolinkurl{https://arxiv.org/abs/#1}}}

\bibitem[{{Bao} {et~al.}(2022){Bao}, {Brotherton}, {Du}, {McLane}, {Zastrocky},
  {Olson}, {Fang}, {Zhai}, {Huang}, {Wang}, {Zhao}, {Li}, {Yang}, {Chen},
  {Liu}, {Yao}, {Peng}, {Guo}, {Songsheng}, {Li}, {Jiang}, {Kasper}, {Chick},
  {Nguyen}, {Maithil}, {Kobulnicky}, {Dale}, {Hand}, {Adelman}, {Carter},
  {Murphree}, {Oeur}, {Schonsberg}, {Roth}, {Winkler}, {Marziani}, {D'Onofrio},
  {Hu}, {Xiao}, {Xue}, {Czerny}, {Aceituno}, {Ho}, {Bai}, {Wang}, \& {MAHA
  Collaboration}}]{Bao_etal_2022}
{Bao}, D.-W., {Brotherton}, M.~S., {Du}, P., {et~al.} 2022, \apjs, 262, 14,
  \dodoi{10.3847/1538-4365/ac7beb}

\bibitem[{{Bentz} {et~al.}(2009){Bentz}, {Peterson}, {Netzer}, {Pogge}, \&
  {Vestergaard}}]{Bentz2009THEAGNs}
{Bentz}, M.~C., {Peterson}, B.~M., {Netzer}, H., {Pogge}, R.~W., \&
  {Vestergaard}, M. 2009, \apj, 697, 160, \dodoi{10.1088/0004-637X/697/1/160}

\bibitem[{{Blandford} \& {McKee}(1982)}]{Blandford1982REVERBERATIONQUASARS}
{Blandford}, R.~D., \& {McKee}, C.~F. 1982, \apj, 255, 419,
  \dodoi{10.1086/159843}

\bibitem[{{Brunthaler} {et~al.}(2005){Brunthaler}, {Falcke}, {Bower}, {Aller},
  {Aller}, \& {Ter{\"a}sranta}}]{Brunthaler2005TheIIIZw2:}
{Brunthaler}, A., {Falcke}, H., {Bower}, G.~C., {et~al.} 2005, \aap, 435, 497,
  \dodoi{10.1051/0004-6361:20042427}

\bibitem[{{Brunthaler} {et~al.}(2000){Brunthaler}, {Falcke}, {Bower}, {Aller},
  {Aller}, {Ter{\"a}sranta}, {Lobanov}, {Krichbaum}, \&
  {Patnaik}}]{Brunthaler2000IIIGalaxy}
---. 2000, \aap, 357, L45, \dodoi{10.48550/arXiv.astro-ph/0004256}

\bibitem[{{Chen} \& {Halpern}(1989)}]{Chen1989STRUCTURE102B}
{Chen}, K., \& {Halpern}, J.~P. 1989, \apj, 344, 115, \dodoi{10.1086/167782}

\bibitem[{{Chen} {et~al.}(1989){Chen}, {Halpern}, \&
  {Filippenko}}]{Chen1989Kinematic102B}
{Chen}, K., {Halpern}, J.~P., \& {Filippenko}, A.~V. 1989, \apj, 339, 742,
  \dodoi{10.1086/167332}

\bibitem[{{Collin} {et~al.}(2006){Collin}, {Kawaguchi}, {Peterson}, \&
  {Vestergaard}}]{Collin2006SystematicInclination}
{Collin}, S., {Kawaguchi}, T., {Peterson}, B.~M., \& {Vestergaard}, M. 2006,
  \aap, 456, 75, \dodoi{10.1051/0004-6361:20064878}

\bibitem[{{Dalla Bont{\`a}} \& {Peterson}(2022)}]{Dallabont2022A}
{Dalla Bont{\`a}}, E., \& {Peterson}, B.~M. 2022, Astronomische Nachrichten,
  343, e210070, \dodoi{10.1002/asna.20210070}

\bibitem[{{Du} \& {Wang}(2019)}]{Du2019TheNuclei}
{Du}, P., \& {Wang}, J.-M. 2019, \apj, 886, 42,
  \dodoi{10.3847/1538-4357/ab4908}

\bibitem[{{Du} {et~al.}(2015){Du}, {Hu}, {Lu}, {Huang}, {Cheng}, {Qiu}, {Li},
  {Zhang}, {Fan}, {Bai}, {Bian}, {Yuan}, {Kaspi}, {Ho}, {Netzer}, {Wang}, \&
  {SEAMBH Collaboration}}]{Du2015SUPERMASSIVEACCRETION}
{Du}, P., {Hu}, C., {Lu}, K.-X., {et~al.} 2015, \apj, 806, 22,
  \dodoi{10.1088/0004-637X/806/1/22}

\bibitem[{Du {et~al.}(2018)Du, Brotherton, Wang, Huang, Hu, Kasper, Chick,
  Nguyen, Maithil, Hand, Li, Ho, Bai, Bian, \& Wang}]{Du2018Mapping}
Du, P., Brotherton, M.~S., Wang, K., {et~al.} 2018, \apj, 869, 142,
  \dodoi{10.3847/1538-4357/aaed2c}

\bibitem[{Eracleous \& Halpern(2003)}]{Eracleous2003CompletionNuclei}
Eracleous, M., \& Halpern, J.~P. 2003, \apj, 599, 886, \dodoi{10.1086/379540}

\bibitem[{{Eracleous} {et~al.}(2009){Eracleous}, {Lewis}, \&
  {Flohic}}]{Eracleous2009Double-peakedNuclei}
{Eracleous}, M., {Lewis}, K.~T., \& {Flohic}, H. M.~L.~G. 2009, \nar, 53, 133,
  \dodoi{10.1016/j.newar.2009.07.005}

\bibitem[{Eracleous {et~al.}(1995)Eracleous, Livio, Halpern, \&
  Storchi-Bergmann}]{Eracleous1995EllipticalNuclei}
Eracleous, M., Livio, M., Halpern, J.~P., \& Storchi-Bergmann, T. 1995, \apj,
  438, 610, \dodoi{10.1086/175104}

\bibitem[{{Gonzalez} {et~al.}(2018){Gonzalez}, {Waddell}, \&
  {Gallo}}]{Gonzalez2017The2}
{Gonzalez}, A.~G., {Waddell}, S.~G.~H., \& {Gallo}, L.~C. 2018, \mnras, 475,
  128, \dodoi{10.1093/mnras/stx3146}

\bibitem[{Gonz{\'{a}}lez-Mart{\'{i}}n
  {et~al.}(2009)Gonz{\'{a}}lez-Mart{\'{i}}n, Rodr{\'{i}}guez-Espinosa,
  D{\'{i}}az-Santos, Packham, Alonso-Herrero, Esquej, Ramos~Almeida, Mason, \&
  Telesco}]{Gonzalez-Martin2009DustPipeline}
Gonz{\'{a}}lez-Mart{\'{i}}n, O., Rodr{\'{i}}guez-Espinosa, J.~M.,
  D{\'{i}}az-Santos, T., {et~al.} 2009, \aap, 553,
  \dodoi{10.1051/0004-6361/201220382}

\bibitem[{Grier {et~al.}(2017)Grier, Pancoast, Barth, Fausnaugh, Brewer, Treu,
  \& Peterson}]{Grier2017TheCampaign}
Grier, C.~J., Pancoast, A., Barth, A.~J., {et~al.} 2017, \apj, 849, 146,
  \dodoi{10.3847/1538-4357/aa901b}

\bibitem[{Grier {et~al.}(2012)Grier, Peterson, Pogge, Denney, Bentz, Martini,
  Sergeev, Kaspi, Minezaki, Zu, Kochanek, Siverd, Shappee, Stanek, Salvo,
  Beatty, Bird, Bord, Borman, Che, Chen, Cohen, Dietrich, Doroshenko, Drake,
  Efimov, Free, Ginsburg, Henderson, King, Koshida, Mogren, Molina, Mosquera,
  Nazarov, Okhmat, Pejcha, Rafter, Shields, Skowron, Szczygiel, Valluri, \&
  Van~Saders}]{Grier2012ReverberationGalaxies}
Grier, C.~J., Peterson, B.~M., Pogge, R.~W., {et~al.} 2012, \apj, 755, 60,
  \dodoi{10.1088/0004-637X/755/1/60}

\bibitem[{Grier {et~al.}(2013)Grier, Peterson, Horne, Bentz, Pogge, Denney,
  De~Rosa, Martini, Kochanek, Zu, Shappee, Siverd, Beatty, Sergeev, Kaspi,
  Salvo, Bird, Bord, Borman, Che, Chen, Cohen, Dietrich, Doroshenko, Efimov,
  Free, Ginsburg, Henderson, King, Mogren, Molina, Mosquera, Nazarov, Okhmat,
  Pejcha, Rafter, Shields, Skowron, Szczygiel, Valluri, \&
  Van~Saders}]{Grier2013TheMaps}
Grier, C.~J., Peterson, B.~M., Horne, K., {et~al.} 2013, \apj, 764, 47,
  \dodoi{10.1088/0004-637X/764/1/47}

\bibitem[{Hung {et~al.}(2020)Hung, Foley, Ramirez-Ruiz, Dai, Auchettl,
  Kilpatrick, Mockler, Brown, Coulter, Dimitriadis, Holoien, Law-Smith, Piro,
  Rest, Rojas-Bravo, \& Siebert}]{HungDouble-peakedEvent}
Hung, T., Foley, R.~J., Ramirez-Ruiz, E., {et~al.} 2020, \apj, 903, 31,
  \dodoi{10.3847/1538-4357/ABB606}

\bibitem[{Kaspi {et~al.}(2000)Kaspi, Smith, Netzer, Maoz, Jannuzi, \&
  Giveon}]{Kaspi2000ReverberationNuclei}
Kaspi, S., Smith, P.~S., Netzer, H., {et~al.} 2000, \apj, 533, 631,
  \dodoi{10.1086/308704}

\bibitem[{Kishimoto {et~al.}(2011)Kishimoto, H{\"{o}}nig, Antonucci, Barvainis,
  Kotani, Tristram, Weigelt, \& Levin}]{Kishimoto2011TheInterferometer}
Kishimoto, M., H{\"{o}}nig, S.~F., Antonucci, R., {et~al.} 2011, \aap, 527,
  A121, \dodoi{10.1051/0004-6361/201016054}

\bibitem[{{Koshida} {et~al.}(2014){Koshida}, {Minezaki}, {Yoshii}, {Kobayashi},
  {Sakata}, {Sugawara}, {Enya}, {Suganuma}, {Tomita}, {Aoki}, \&
  {Peterson}}]{Koshida2014}
{Koshida}, S., {Minezaki}, T., {Yoshii}, Y., {et~al.} 2014, \apj, 788, 159,
  \dodoi{10.1088/0004-637X/788/2/159}

\bibitem[{Liao {et~al.}(2016)Liao, Xin, Fan, Weng, Li, Chen, \&
  Fan}]{Liao2016DiscoveryVariability}
Liao, N.-H., Xin, Y.-L., Fan, X.-L., {et~al.} 2016, \apjs, 226, 17,
  \dodoi{10.3847/0067-0049/226/2/17}

\bibitem[{Marinello {et~al.}(2016)Marinello, Rodr{\'{i}}guez-Ardila,
  Garcia-Rissmann, Sigut, \& Pradhan}]{Marinello2016}
Marinello, M., Rodr{\'{i}}guez-Ardila, A., Garcia-Rissmann, A., Sigut, T.
  A.~A., \& Pradhan, A.~K. 2016, \apj, 820, 116,
  \dodoi{10.3847/0004-637x/820/2/116}

\bibitem[{{Marinello} {et~al.}(2020){Marinello}, {Rodr{\'\i}guez-Ardila},
  {Marziani}, {Sigut}, \& {Pradhan}}]{Marinello2020Panchromatic1092}
{Marinello}, M., {Rodr{\'\i}guez-Ardila}, A., {Marziani}, P., {Sigut}, A., \&
  {Pradhan}, A. 2020, \mnras, 494, 4187, \dodoi{10.1093/mnras/staa934}

\bibitem[{{Mart{\'\i}nez-Aldama} {et~al.}(2015){Mart{\'\i}nez-Aldama},
  {Dultzin}, {Marziani}, {Sulentic}, {Bressan}, {Chen}, \&
  {Stirpe}}]{LoliMartinez-Aldama2015}
{Mart{\'\i}nez-Aldama}, M.~L., {Dultzin}, D., {Marziani}, P., {et~al.} 2015,
  \apjs, 217, 3, \dodoi{10.1088/0067-0049/217/1/3}

\bibitem[{Mason {et~al.}(2015)Mason, Rodr{\'{i}}guez-Ardila, Martins, Riffel,
  Mart{\'{i}}n, Almeida, Dutra, Ho, Thanjavur, Flohic, Alonso-Herrero, Lira,
  McDermid, Riffel, Schiavon, Winge, Hoenig, \& Perlman}]{Mason2015THEGALAXIES}
Mason, R.~E., Rodr{\'{i}}guez-Ardila, A., Martins, L., {et~al.} 2015, \apjs,
  217, 13, \dodoi{10.1088/0067-0049/217/1/13}

\bibitem[{Matsuoka {et~al.}(2008)Matsuoka, Kawara, \&
  Oyabu}]{Matsuoka2008LowIonizationLines}
Matsuoka, Y., Kawara, K., \& Oyabu, S. 2008, \apj, 673, 62,
  \dodoi{10.1086/524193}

\bibitem[{{Nenkova} {et~al.}(2008){Nenkova}, {Sirocky}, {Nikutta},
  {Ivezi{\'c}}, \& {Elitzur}}]{Nenkova2008}
{Nenkova}, M., {Sirocky}, M.~M., {Nikutta}, R., {Ivezi{\'c}}, {\v{Z}}., \&
  {Elitzur}, M. 2008, \apj, 685, 160, \dodoi{10.1086/590483}

\bibitem[{{Newville} {et~al.}(2014){Newville}, {Stensitzki}, {Allen}, \&
  {Ingargiola}}]{Newville2014LMFIT:Python}
{Newville}, M., {Stensitzki}, T., {Allen}, D.~B., \& {Ingargiola}, A. 2014,
  {LMFIT: Non-Linear Least-Square Minimization and Curve-Fitting for Python},
  0.8.0, Zenodo,  Zenodo, \dodoi{10.5281/zenodo.11813}

\bibitem[{Pancoast {et~al.}(2014)Pancoast, Brewer, Treu, Park, Barth, Bentz, \&
  Woo}]{pancoast14}
Pancoast, A., Brewer, B.~J., Treu, T., {et~al.} 2014, Monthly Notices of the
  Royal Astronomical Society, 445, 3073, \dodoi{10.1093/mnras/stu1419}

\bibitem[{{Panda}(2021)}]{2021POBeo.100..333P}
{Panda}, S. 2021, in XIX Serbian Astronomical Conference, Vol. 100, 333--338

\bibitem[{{Panda} {et~al.}(2020){Panda}, {Mart{\'\i}nez-Aldama}, {Marinello},
  {Czerny}, {Marziani}, \& {Dultzin}}]{Panda2020OpticalModelling}
{Panda}, S., {Mart{\'\i}nez-Aldama}, M.~L., {Marinello}, M., {et~al.} 2020,
  \apj, 902, 76, \dodoi{10.3847/1538-4357/abb5b8}

\bibitem[{{Panda} {et~al.}(2019){Panda}, {Marziani}, \&
  {Czerny}}]{Pandaetal2019}
{Panda}, S., {Marziani}, P., \& {Czerny}, B. 2019, \apj, 882, 79,
  \dodoi{10.3847/1538-4357/ab3292}

\bibitem[{{Pecaut} \& {Mamajek}(2013)}]{pecaut2013intrinsic}
{Pecaut}, M.~J., \& {Mamajek}, E.~E. 2013, \apjs, 208, 9,
  \dodoi{10.1088/0067-0049/208/1/9}

\bibitem[{Peterson \& Horne(2004)}]{Peterson2004ReverberationNuclei}
Peterson, B.~M., \& Horne, K. 2004, Astronomische Nachrichten, 325, 248,
  \dodoi{10.1002/asna.200310207}

\bibitem[{{Peterson} {et~al.}(2004){Peterson}, {Ferrarese}, {Gilbert}, {Kaspi},
  {Malkan}, {Maoz}, {Merritt}, {Netzer}, {Onken}, {Pogge}, {Vestergaard}, \&
  {Wandel}}]{Peterson2004BlackMeasurements}
{Peterson}, B.~M., {Ferrarese}, L., {Gilbert}, K.~M., {et~al.} 2004, \apj, 613,
  682, \dodoi{10.1086/423269}

\bibitem[{{Popovi{\'c}} {et~al.}(2003){Popovi{\'c}}, {Mediavilla}, {Bon},
  {Stani{\'c}}, \& {Kubi{\v{c}}ela}}]{Popovic2003TheVariability}
{Popovi{\'c}}, L.~{\v{C}}., {Mediavilla}, E.~G., {Bon}, E., {Stani{\'c}}, N.,
  \& {Kubi{\v{c}}ela}, A. 2003, \apj, 599, 185, \dodoi{10.1086/379277}

\bibitem[{{Reunanen} {et~al.}(2002){Reunanen}, {Kotilainen}, \&
  {Prieto}}]{2002reunanen}
{Reunanen}, J., {Kotilainen}, J.~K., \& {Prieto}, M.~A. 2002, \mnras, 331, 154,
  \dodoi{10.1046/j.1365-8711.2002.05181.x}

\bibitem[{Riffel {et~al.}(2006)Riffel, Rodr{\'{i}}guez-Ardila, \&
  Pastoriza}]{Riffel2006}
Riffel, R., Rodr{\'{i}}guez-Ardila, A., \& Pastoriza, M.~G. 2006, \aap, 457,
  61, \dodoi{10.1051/0004-6361:20065291}

\bibitem[{Rodriguez‐Ardila {et~al.}(2000)Rodriguez‐Ardila, Pastoriza, \&
  Donzelli}]{RodriguezArdila2000VisibleGalaxies}
Rodriguez‐Ardila, A., Pastoriza, M.~G., \& Donzelli, C.~J. 2000, \apjs, 126,
  63, \dodoi{10.1086/313293}

\bibitem[{Schimoia {et~al.}(2012)Schimoia, Storchi-Bergmann, Nemmen, Winge, \&
  Eracleous}]{Schimoia2012SHORT1097}
Schimoia, J.~S., Storchi-Bergmann, T., Nemmen, R.~S., Winge, C., \& Eracleous,
  M. 2012, \apj, 748, 145, \dodoi{10.1088/0004-637X/748/2/145}

\bibitem[{Schimoia {et~al.}(2017)Schimoia, Storchi-Bergmann, Winge, Nemmen,
  Eracleous, Schimoia, Storchi-Bergmann, Winge, Nemmen, \&
  Eracleous}]{Schimoia2017Evolution7213}
Schimoia, J.~S., Storchi-Bergmann, T., Winge, C., {et~al.} 2017, MNRAS, 472,
  2170, \dodoi{10.1093/MNRAS/STX2107}

\bibitem[{Sergeev(2020)}]{sergeev20}
Sergeev, S.~G. 2020, Monthly Notices of the Royal Astronomical Society, 495,
  971, \dodoi{10.1093/mnras/staa1210}

\bibitem[{Shapovalova {et~al.}(2010)Shapovalova, Popovi{\'{c}}, Burenkov,
  Chavushyan, Ili{\'{c}}, Kova{\v{c}}evi{\'{c}}, Bochkarev, \&
  Le{\'{o}}n-Tavares}]{Shapovalova2010Long-termProfiles}
Shapovalova, A.~I., Popovi{\'{c}}, L.~C., Burenkov, A.~N., {et~al.} 2010, \aap,
  509, A106, \dodoi{10.1051/0004-6361/200912311}

\bibitem[{Shapovalova {et~al.}(2008)Shapovalova, Popovi{\'{c}}, Collin,
  Burenkov, Chavushyan, Bochkarev, Ben{\'{i}}tez, Dultzin,
  Kova{\v{c}}evi{\'{c}}, Borisov, Carrasco, Le{\'{o}}n-Tavares, Mercado,
  Valdes, Vlasuyk, \& Zhdanova}]{Shapovalova2008Long-termCorrelations}
Shapovalova, A.~I., Popovi{\'{c}}, L.~C., Collin, S., {et~al.} 2008, \apjs,
  486, 99, \dodoi{10.1051/0004-6361:20079111}

\bibitem[{Shapovalova {et~al.}(2012)Shapovalova, Popovi{\'{c}}, Burenkov,
  Chavushyan, Ili{\'{c}}, Kova{\v{c}}evi{\'{c}}, Kollatschny,
  Kova{\v{c}}evi{\'{c}}, Bochkarev, Valdes, Torrealba, Le{\'{o}}n-Tavares,
  Mercado, Ben{\'{i}}tez, Carrasco, Dultzin, \& Fuente}]{Shapovalova2012}
Shapovalova, A.~I., Popovi{\'{c}}, L.~C., Burenkov, A.~N., {et~al.} 2012, \aap,
  202, 10, \dodoi{10.1088/0067-0049/202/1/10}

\bibitem[{Skrutskie {et~al.}(2006)Skrutskie, Cutri, Stiening, Weinberg,
  Schneider, Carpenter, Beichman, Capps, Chester, Elias,
  {et~al.}}]{skrutskie2006two}
Skrutskie, M., Cutri, R., Stiening, R., {et~al.} 2006, \aj, 131, 1163

\bibitem[{Storchi-Bergmann {et~al.}(2017)Storchi-Bergmann, Schimoia, Peterson,
  Elvis, Denney, Eracleous, \&
  Nemmen}]{Storchi-Bergmann2017Double-PeakedNuclei}
Storchi-Bergmann, T., Schimoia, J.~S., Peterson, B.~M., {et~al.} 2017, \apj,
  835, 236, \dodoi{10.3847/1538-4357/835/2/236}

\bibitem[{Strateva {et~al.}(2003)Strateva, Strauss, Hao, Schlegel, Hall, Gunn,
  Li, Ivezi, Richards, Zakamska, Voges, Anderson, Lupton, Schneider, Brinkmann,
  \& Nichol}]{strateva}
Strateva, I.~V., Strauss, M.~A., Hao, L., {et~al.} 2003, \aj, 126, 1720,
  \dodoi{10.1086/378367}

\bibitem[{{Zhang} {et~al.}(2019){Zhang}, {Zhou}, {Shi}, {Ji}, {Jiang}, {Pan},
  {Sheng}, {Sun}, \& {Zhong}}]{zhang2019}
{Zhang}, S., {Zhou}, H., {Shi}, X., {et~al.} 2019, \apj, 877, 33,
  \dodoi{10.3847/1538-4357/ab1aa3}

\end{thebibliography}
\bibliographystyle{aasjournal}



\end{document}